\documentclass[12pt,a4paper,reqno,tbtags,oneside]{amsproc}
\usepackage{amssymb}
\usepackage{amsmath}
\usepackage{color}
\usepackage[colorlinks]{hyperref}

\input tcilatex

\begin{document}
\title{Gurevich-Zybin system}
\author{Maxim V.Pavlov}
\address{Institute of Mathematics, Academia Sinica, Taipei 11529, Taiwan\\
e-mail: m.v.pavlov@lboro.ac.uk}

\begin{abstract}
We present three different linearizable extensions of the Gurevich-Zybin
system. Their general solutions are found by reciprocal transformations.

In this paper we rewrite the Gurevich-Zybin system as a Monge-Ampere
equation. By application of reciprocal transformation this equation is
linearized. Infinitely many local Hamiltonian structures, local Lagrangian
representations, local conservation laws and local commuting flows are
found. Moreover, all commuting flows can be written as Monge-Ampere
equations similar to the Gurevich-Zybin system.

The Gurevich-Zybin system describes the formation of a large scale
structures in the Universe. The second harmonic wave generation is known in
nonlinear optics. In this paper we prove that the Gurevich-Zybin system is
equivalent to a degenerate case of the second harmonic generation. Thus, the
Gurevich-Zybin system is recognized as a degenerate first negative flow of
two-component Harry Dym hierarchy up to two Miura type transformations. A
reciprocal transformation between the Gurevich-Zybin system and degenerate
case of the second harmonic generation system is found. A new solution for
the second harmonic generation is presented in implicit form.
\end{abstract}

\maketitle

\vspace{1cm}

to the memory of Professor Andrea Donato (Messina University)

\vspace{1cm}

\tableofcontents
\section{Introduction}

The invisible nondissipative dark matter plays a decisive role in the
formation of a large scale structures in the Universe: galaxies, clusters of
galaxies, superclusters. Corresponding nonlinear dynamics can be described
(see \cite{gz}) by following hydrodynamic-like system%
\begin{equation}
\rho _{t}+\nabla (\rho \mathbf{u})=0\text{, \ \ \ \ }u_{t}+(\mathbf{u}\nabla
)u+\nabla \Phi =0\text{, \ \ \ }\Delta \Phi =\rho ,  \label{1}
\end{equation}%
where the first two equations are usual hydrodynamic equations (the
continuity equation and the Euler equation, respectively), but the third
equation is the famous Poisson equation. At least for the first time this
system was derived by J.H. Jeans (see \cite{jeans} and also \cite{zeld}) for
a description of instabilities of a homogeneous distribution of a matter.

Such dynamics of dissipationless gravitating gas is a special limit ($%
\varepsilon \rightarrow 0$) of another system (another sign of $\rho $ is
unessential)%
\begin{equation*}
\rho _{t}+\nabla (\rho \mathbf{u})=0\text{, \ \ \ \ }u_{t}+(\mathbf{u}\nabla
)u+\nabla \Phi =0\text{, \ \ \ }\Delta \Phi =\varepsilon e^{\Phi }-\rho
\end{equation*}%
describing fully nonlinear flows in a two temperature unmagnetized
collisionless plasma in dimensionless variables (nonlinear ion-acoustic
waves; see for instance \cite{sagdeev}).

The main advantage of the Jeans theory is a reckoning of two factors: a
gravity attracting a matter in separate lumps and clots; a pressure
decreasing an inhomogeneity of a matter in the Universe.

Recently a new achievement in the investigation of the system
(\ref{1}) was made (see \cite{gz}) in Cosmology. The nonlinear
one-dimensional dynamics of a dark matter is described by the
equations \cite{gz}
\begin{equation}
u_{t}+uu_{x}+\upsilon =0\text{, \ \ \ \ \ \ \ \ \ }\upsilon _{t}+u\upsilon
_{x}=0,  \label{3}
\end{equation}%
where $\rho =\upsilon _{x}$, \ $\upsilon =\Phi _{x}$. The analysis of
equations (\ref{3}) in a multimode form demonstrates the transition from the
hydrodynamic to the equilibrium kinetic state \cite{gz}. It means, that the
exact solution of the equations (\ref{3}) describes a fundamental physical
process (see \cite{gz} again).

\qquad It is amazing that the \textit{inhomogeneous} hydrodynamic type
system (\ref{3}) can be integrated, up to the first singularity, by the
Hodograph Method (see \cite{gz}). For this reason further we will call the
system (\ref{3}) the Gurevich-Zybin system emphasizing that the
one-dimensional reduction (\ref{3}) of the system (\ref{1}) is integrable.

Here we give the general solution by the method of \textit{Reciprocal
Transformations}. Moreover, we present three different linearizable
extensions of this system (\ref{3}) with their general solutions given by
corresponding reciprocal transformations. Actually these reciprocal
transformations have clear pure mathematical (hodograph method) and physical
(transition from Euler to Lagrange variables) interpretations. In the next
section we present three linearizable extensions of the Gurevich-Zybin
systems with their general solutions. In second section relationship between
two-component generalization of the Hunter-Saxton equation and the
Gurevich-Zybin system is established. In the third section the
Gurevich-Zybin system is rewritten as a Monge-Ampere equation (following the
approach developed by Andrea Donato). In the fourth section a bi-Hamiltonian
structure of the Gurevich-Zybin system is found (following the approach
developed by Yavuz Nutku). In the fifth section, by the application of a
reciprocal transformation, a simplest recursion operator is constructed.
Infinitely many local conservation laws, local commuting flows, local
Lagrangians and local Hamiltonians are found. Moreover, all commuting flows
are Monge-Ampere equations. Thus, the Gurevich-Zybin system is a member of
an integrable hierarchy of Monge-Ampere equations. In sixth section a
bi-Hamiltonian formulation for the Gurevich-Zybin system is given in a
canonical form. The Gurevich-Zybin system is recognized as a first negative
flow of two-component Harry Dym hierarchy. In seventh section Miura type and
reciprocal transformations between the Gurevich-Zybin system and
Kaup-Boussinesq hierarchy are given. In eighth section we finally prove that
the Gurevich-Zybin system is equivalent to a degenerate case of the second
harmonic generation system up to above-mentioned transformations. A new
solutions of the second harmonic generation system is found. In conclusion
we discuss about sort of integrable problems belonging some different
hierarchies of integrable equations.

\section{General Solution}

The Gurevich-Zybin system (\ref{1}) in one dimensional case precisely has a
form%
\begin{equation}
\rho _{t}+\partial _{x}(\rho u)=0\text{, \ \ \ }u_{t}+uu_{x}+\Phi _{x}=0%
\text{, \ \ \ \ }\Phi _{xx}=\rho .  \label{2}
\end{equation}%
This system can be generalized \textit{at least} in three different forms:%
\begin{eqnarray}
\rho _{t}+\partial _{x}(\rho u) &=&0\text{, \ \ \ }u_{t}+uu_{x}+\mu ^{\prime
\prime \prime }(\Phi _{x})=0\text{, \ \ \ \ }\Phi _{xx}=\rho ,  \label{4} \\
\rho _{t}+\partial _{x}(\rho u) &=&0\text{, \ \ \ }u_{t}+uu_{x}+\Phi _{x}=0%
\text{, \ \ \ \ \ \ \ \ \ \ }\Phi _{xx}=B(\rho ),  \label{5} \\
\rho _{t}+\partial _{x}(\rho u) &=&0\text{, \ \ \ }u_{t}+uu_{x}+\mu ^{\prime
\prime \prime }(\Phi _{x})=0\text{, \ \ \ \ }\partial _{x}C(\Phi _{x})=\rho ,
\label{6}
\end{eqnarray}%
where $\mu (z)$, $B(\rho)$ and $C(z)$ are arbitrary functions. It
is interesting that is not obvious that the system (\ref{4}) is
equivalent to the Gurevich-Zybin system (\ref{2}). Indeed, the
system (\ref{4}) written
like (\ref{3})%
\begin{equation}
z_{t}+uz_{x}=0\text{, \ \ \ \ }u_{t}+uu_{x}+\mu ^{\prime \prime \prime }(z)=0
\label{ex}
\end{equation}%
is exactly (\ref{3}) up to point transformation $\upsilon =\mu ^{\prime
\prime \prime }(z)$. One can introduce reciprocal transformation%
\begin{equation}
dz=\rho dx-\rho udt\text{, \ \ \ \ }d\tau =dt.  \label{7}
\end{equation}%
Then $\partial _{x}=\rho \partial _{z}$ and $\partial _{t}=\partial _{\tau
}-\rho u\partial _{z}$. Thus, the system (\ref{4}) has \ the form%
\begin{equation}
(\frac{1}{\rho })_{\tau }=u_{z}\text{, \ \ \ \ \ }u_{\tau }=-\mu ^{\prime
\prime \prime }(z)\text{,}  \label{8}
\end{equation}%
in new variables, where $z=\Phi _{x}$. Thus, the general solution of the
system (\ref{8}) is%
\begin{equation*}
u=-\mu ^{\prime \prime \prime }(z)\tau +D^{\prime }(z)\text{, \ \ \ }\frac{1%
}{\rho }=-\mu ^{\prime \prime \prime \prime }(z)\tau ^{2}/2+D^{\prime \prime
}(z)\tau +E^{\prime \prime }(z),
\end{equation*}%
where $D(z)$ and $E(z)$ are arbitrary functions. Finally the general
solution of the system (\ref{4}) can be given in implicit form%
\begin{eqnarray}
u &=&-\mu ^{\prime \prime \prime }(z)t+D^{\prime }(z)\text{, \ \ \ }\rho
=[-\mu ^{\prime \prime \prime \prime }(z)t^{2}/2+D^{\prime \prime
}(z)t+E^{\prime \prime }(z)]^{-1}\text{,}  \notag \\
x &=&-\mu ^{\prime \prime \prime }(z)t^{2}/2+D^{\prime }(z)t+E^{\prime }(z),
\label{9} \\
\Phi &=&(\mu ^{\prime \prime }(z)-z\mu ^{\prime \prime \prime
}(z))t^{2}/2+(zD^{\prime }(z)-D(z))t+zE^{\prime }(z)-E(z),  \notag
\end{eqnarray}%
where $z$ is a parameter here.

Above reciprocal transformation applied to the system (\ref{5}) yields%
\begin{equation*}
(\frac{1}{\rho })_{\tau }=u_{z}\text{, \ \ \ \ \ }u_{\tau }=-\upsilon \text{%
, \ \ \ \ }\upsilon _{z}=B(\rho )/\rho ,
\end{equation*}%
where $\upsilon =\Phi _{x}$. If function $\rho $ can be explicitly expressed
from algebraic equation%
\begin{equation*}
\tau =G(z)-\overset{\rho }{\int }\frac{d\theta }{\theta ^{2}\sqrt{2\int
B(\theta )\theta ^{-3}d\theta -F(z)}},
\end{equation*}%
where $F(z)$ and $G(z)$ are arbitrary functions, then the general
solution of the system (\ref{5}) can be obtained. For instance, if
$B(\rho )=\rho $, then a general solution is already given by
(\ref{9}) (remember that in such sub-case systems (\ref{4}) and
(\ref{5}) coincide if $\mu (z)=z^{4}/24$); in simplest perturbed
case $B(\rho )=\rho +\delta /\rho $ ($\delta =\func{const}
$) the general solution is expressed via Weiershtrass elliptic functions%
\begin{equation*}
\rho =\frac{\delta }{6}[\wp (\tau -G(z)\text{, }\frac{\delta }{3}\text{, }%
\frac{\delta ^{2}}{36}F(z))]^{-1}\text{, \ \ \ }u_{z}=\frac{6}{\delta }\wp
^{\prime }\text{, \ \ \ }\upsilon _{z}=1-\frac{36}{\delta }\wp ^{2}.
\end{equation*}

The reciprocal transformation (\ref{7}) applied to the system (\ref{6})
yields%
\begin{equation*}
(\frac{1}{\rho })_{\tau }=u_{z}\text{, \ \ \ \ \ }u_{\tau }=-\mu ^{\prime
\prime \prime }(\upsilon )\text{, \ \ \ \ }C(\upsilon )=z\text{,}
\end{equation*}%
where $\upsilon =\Phi _{x}$. Thus, the general solution of this system is%
\begin{eqnarray*}
\upsilon &=&V(z)\text{, \ \ \ \ \ }u=D^{\prime }(z)-\mu ^{\prime \prime
\prime }(\upsilon )\tau \text{,} \\
\rho &=&[E^{\prime \prime }(z)+D^{\prime \prime }(z)\tau -\frac{\mu ^{\prime
\prime \prime \prime }(\upsilon )}{2C^{\prime }(\upsilon )}\tau ^{2}]^{-1}%
\text{,} \\
x &=&E^{\prime }(z)+D^{\prime }(z)\tau -\mu ^{\prime \prime \prime
}(\upsilon )\tau ^{2}/2,
\end{eqnarray*}%
where $D(z)$, $E(z)$ are arbitrary functions and $V(z)$ is inverse function
to $C(\upsilon )$. Finally the general solution of the system (\ref{6}) can
be given in implicit form%
\begin{eqnarray*}
\upsilon &=&V(z)\text{, \ \ \ \ \ }u=D^{\prime }(z)-\mu ^{\prime \prime
\prime }(\upsilon )t\text{,} \\
\rho &=&[E^{\prime \prime }(z)+D^{\prime \prime }(z)t-\frac{\mu ^{\prime
\prime \prime \prime }(\upsilon )}{2C^{\prime }(\upsilon )}t^{2}]^{-1}\text{,%
} \\
x &=&E^{\prime }(z)+D^{\prime }(z)t-\mu ^{\prime \prime \prime }(\upsilon
)t^{2}/2, \\
\Phi &=&zE^{\prime }(z)-E(z)+(zD^{\prime }(z)-D(z))t+[G(\upsilon
)-zG^{\prime }(\upsilon )]t^{2}/2,
\end{eqnarray*}%
where $G^{\prime }=\mu ^{\prime \prime \prime }C^{\prime }$ and $z$ is a
parameter here. Also the system (\ref{6}) can be written in hydrodynamic
like form%
\begin{equation}
\rho _{t}+\partial _{x}(\rho u)=0\text{, \ \ \ \ }u_{t}+uu_{x}+\frac{1}{\rho
}\partial _{x}P=0,  \label{10}
\end{equation}%
where the pressure $P$ is a nonlocal function of the density $\rho $%
\begin{equation*}
P=P(V(\partial _{x}^{-1}\rho )).
\end{equation*}%
In particular case $C(\upsilon )=\upsilon $ the system (\ref{6}) coincides
with the system (\ref{4}); the system (\ref{6}) was written in form (\ref{10}%
) in \cite{das} for particular case $\mu (\upsilon )=\upsilon ^{4}/24$ and $%
C(\upsilon )=\upsilon $.

\section{Two-component generalization of the Calogero equation}

The simplest two-component linearizable generalization%
\begin{equation}
\eta _{t}+\partial _{x}(\eta u)=0\text{, \ \ \ \ \ \ }u_{xt}+uu_{xx}+\Psi
(\eta \text{, }u_{x})=0  \label{11}
\end{equation}%
of the Calogero equation (see \cite{fc})%
\begin{equation*}
u_{xt}=uu_{xx}+R(u_{x})
\end{equation*}%
was presented (functions $\Psi (a,b)$ and $R(c)$ are arbitrary here) in \cite%
{mpl}. Some particular cases of the Calogero equation like the
Hunter-Saxton equation (i.e. $R(c)=\nu c^{2}$, where $\nu
=\func{const}$; see \cite{hs}) are interested from physical point
of view. The Calogero equation is linearizable by a reciprocal
transformation (see \cite{mpl}). For instance, the Hunter-Saxton
equation is related to the famous Liouville equation by a
reciprocal transformation (see \cite{dp}). Thus, the system
(\ref{11}) is a natural generalization of the Liouville equation
on two-component case up to module of a (invertible) reciprocal
transformation.

The reciprocal transformation%
\begin{equation}
d\zeta =\eta dx-\eta udt\text{, \ \ \ }dy=dt  \label{eta}
\end{equation}%
applied to the system (\ref{11}) yields the \textit{ordinary} differential
equation%
\begin{equation*}
s_{yy}+\Psi (e^{-s}\text{, }s_{y})=0,
\end{equation*}%
where $s=-\ln \eta $, which can be reduced to the first order equation%
\begin{equation}
ada+\Psi (e^{-s}\text{, }a)ds=0,  \label{12}
\end{equation}%
where $a=s_{y}=u_{x}$. Then a general solution can be constructed in two
steps from%
\begin{equation*}
u_{\zeta }=(\frac{1}{\eta })_{y}\text{, \ \ \ }dx=\frac{1}{\eta }d\zeta +udy.
\end{equation*}%
A solution $q(s$, $a)$ of the linear equation%
\begin{equation}
\frac{\partial q}{\partial s}=\frac{\Psi (e^{-s}\text{, }a)}{a}\frac{%
\partial q}{\partial a}  \label{lin}
\end{equation}%
determined by the characteristic equation (\ref{12}) yields extra
conservation law%
\begin{equation*}
\rho _{t}+\partial _{x}(\rho u)=0,
\end{equation*}%
where $\rho (\eta $, $u_{x})=\eta \exp q$. The comparison of the
second equation in (\ref{11}) and the second equation in (\ref{2})
yields another
relationship%
\begin{equation*}
\Psi =\rho +u_{x}^{2}.
\end{equation*}%
Thus, a solution of the \textit{nonlinear} equation (substitute $\Psi $ from
above equation into (\ref{lin}))%
\begin{equation*}
q_{s}=(a+\frac{e^{q-s}}{a})q_{a}
\end{equation*}%
describes a transformation between the Gurevich-Zybin system
(\ref{2}) and two-component generalization of the Hunter-Saxton
equation (\ref{11}).

\textbf{Remark 1}: \textit{Above equation under the substitution}%
\begin{equation*}
q=s+\ln (\frac{n}{2}e^{-3s}-a^{2})
\end{equation*}%
\textit{transforms into well-known inhomogenious Riemann-Monge-Hopf equation}%
\begin{equation*}
n_{y}+nn_{c}=-\frac{2c}{9y^{2}},
\end{equation*}%
\textit{where }$y=e^{-3s}/3$\textit{\ and }$c=a^{2}$\textit{. Its general
solution can be given just in parametric form}%
\begin{equation*}
n=\frac{1}{3}A_{1}(\xi )y^{-2/3}+\frac{2}{3}A_{2}(\xi )y^{-1/3}\text{, \ \ \
\ \ }c=A_{1}(\xi )y^{1/3}+A_{2}(\xi )y^{2/3},
\end{equation*}%
\textit{where }$\xi $\textit{\ is a parameter and }$A_{1}(\xi )$\textit{, }$%
A_{2}(\xi )$\textit{\ are arbitrary functions. However, the general solution
of the equation of the first order depends by \textbf{one} function of a
single variable only. It means, that if for instance }$A_{1}(\xi )\neq
\limfunc{const}$\textit{, then by re-scaling }$A_{1}(\xi )\rightarrow \xi $%
\textit{\ the general solution take a form}%
\begin{equation*}
n=\frac{1}{3}\xi y^{-2/3}+\frac{2}{3}A(\xi )y^{-1/3}\text{, \ \ \ \ \ }c=\xi
y^{1/3}+A(\xi )y^{2/3},
\end{equation*}%
where $A(\xi )$ is an arbitrary function.

In a particular case the substitutions (see \cite{mpl})%
\begin{equation}
\rho =-\frac{1}{2}(u_{x}^{2}+\eta ^{2})\text{, \ \ \ \ \ }\Psi (\eta \text{,
}u_{x})=\frac{1}{2}(u_{x}^{2}-\eta ^{2})  \label{ra}
\end{equation}%
connect the Gurevich-Zybin system (\ref{2}) with two-component
Hunter-Saxton system (see (\ref{11}) and \cite{or}), which is a
\textit{bi-directional} version of the Hunter-Saxton equation. A
general solution of the Gurevich-Zybin system in field variables
$\eta $ and $u$ has the implicit form
with respect to the parameter $\zeta $%
\begin{eqnarray}
\eta &=&[\frac{1}{k^{\prime }(\zeta )}+\frac{1}{4}k^{\prime }(\zeta
)(t-m(\zeta ))^{2}]^{-1},  \notag \\
u &=&\frac{1}{2}tk(\zeta )-\frac{1}{2}\int m(\zeta )dk(\zeta ),  \label{13}
\\
x &=&\int [\frac{1}{k^{\prime }(\zeta )}+\frac{1}{4}k^{\prime }(\zeta
)m^{2}(\zeta )]d\zeta +\frac{1}{4}t^{2}k(\zeta )-\frac{1}{2}t\int m(\zeta
)dk(\zeta ),  \notag
\end{eqnarray}%
where $m(\zeta )$, $k(\zeta )$ are arbitrary functions. One can
substitute above expression for $\eta $ to the first equation in
(\ref{ra}) and compare expressions for ($\rho $, $x$, $u$) from
(\ref{9}) (in this case $\mu ^{\prime \prime \prime }(z)=z$) and
above, then a relationship between the arbitary functions $k$, $m$
and $D$, $E$ will be reconstructed.

\section{Reformulation of the Gurevich-Zybin system as a Monge-Ampere
equation}

A lot of physically motivated nonlinear systems can be written as
Monge-Ampere equations (see \cite{cr}). To this moment we have no unique
method for constructing such relationships. One simple approach was
suggested to author by Andrea Donato (see \cite{do}) at the
\textquotedblleft Lie Group Analysis\textquotedblright\ conference in
Johannesburg (South Africa) at 1996.

The Gurevich-Zybin system in physical field variables (\ref{2}) has four
local conservation laws%
\begin{eqnarray*}
u_{t}+\partial _{x}(u^{2}/2+\Phi ) &=&0\text{, \ \ \ \ \ }\rho _{t}+\partial
_{x}(\rho u)=0\text{,} \\
\partial _{t}(\rho u)+\partial _{x}(\rho u^{2}+\Phi _{x}^{2}/2) &=&0\text{,
\ \ \ \ \ }\partial _{t}(\rho u^{2}-\Phi _{x}^{2})+\partial _{x}(\rho
u^{3})=0
\end{eqnarray*}%
as a consequence of obvious local Hamiltonian structure%
\begin{equation}
\upsilon _{t}=\frac{\delta H_{2}}{\delta u}\text{, \ \ \ \ \ }u_{t}=-\frac{%
\delta H_{2}}{\delta \upsilon },  \label{sev}
\end{equation}%
where the Hamiltonian is $H_{2}=\frac{1}{2}\int [-u^{2}\upsilon
_{x}+\upsilon ^{2}]dx$, the momentum is $H_{1}=\int u\upsilon _{x}dx$, and
two Casimirs are functionals $Q_{1}=\int udx$ and $Q_{2}=\int \rho dx$ of
corresponding Poisson bracket%
\begin{equation}
\{\rho (x)\text{, }u(x^{\prime })\}=\{u(x)\text{, }\rho (x^{\prime
})\}=\delta ^{\prime }(x-x^{\prime }).  \label{eight}
\end{equation}%
The existence of above first three local conservation laws is obvious.
However, the fourth conservation law is not easy to check. Since $\rho =\Phi
_{xx}$, then $\rho u=-\Phi _{xt}$, then $\rho u^{2}+\Phi _{x}^{2}/2=\Phi
_{tt}$, then above fourth conservation law is valid. Eliminating physical
field variables $\rho $ and $u$ from these three equations, the Monge-Ampere
equation is given by%
\begin{equation}
\Phi _{xx}\Phi _{tt}-\Phi _{xt}^{2}=\frac{1}{2}\Phi _{x}^{2}\Phi _{xx}.
\label{mon}
\end{equation}%
In paper \cite{gz} the Gurevich-Zybin system was linearized by a
hodograph method. A general solution has been presented too. Thus,
this Monge-Ampere equation is linearizable and has the general
solution in implicit form (see
the end of section \textbf{2})%
\begin{equation*}
\Phi =-\frac{1}{4}\upsilon ^{2}t^{2}+(\upsilon D^{\prime }(\upsilon
)-D(\upsilon ))t+\upsilon E^{\prime }(\upsilon )-E(\upsilon )\text{, \ \ \ \
\ \ \ }x=-\upsilon t^{2}/2+D^{\prime }(\upsilon )t+E^{\prime }(\upsilon ),
\end{equation*}%
where $D(\upsilon )$ and $E(\upsilon )$ are arbitrary functions, $\upsilon $
is a parameter here.

Since the Gurevich-Zybin system can be written in the form (\ref{4}), (\ref%
{ex}), we shall use an arbitrary value of function $\mu (z)$ in next \textbf{%
two} sections.

\section{Bi-Hamiltonian structure}

The Gurevich-Zybin system (\ref{ex}) has local bi-Hamiltonian structure,
where the first local Hamiltonian structure is (\ref{sev})%
\begin{equation}
z_{t}=\frac{\delta H_{2}}{\delta u}\text{, \ \ \ \ \ }u_{t}=-\frac{\delta
H_{2}}{\delta z},  \label{fir}
\end{equation}%
just three conservation laws are
\begin{eqnarray*}
\partial _{t}\rho +\partial _{x}(\rho u) &=&0, \\
\partial _{t}(\rho u)+\partial _{x}[\rho u^{2}+\mu ^{\prime \prime }(\Phi
_{x})] &=&0, \\
\partial _{t}[\rho u^{2}-2\mu ^{\prime \prime }(\Phi _{x})]+\partial
_{x}(\rho u^{3}) &=&0,
\end{eqnarray*}%
associated with the first Poisson bracket (\ref{eight}). The Hamiltonian is $%
H_{2}=\int [-\rho u^{2}+2\mu ^{\prime \prime }(\Phi _{x})]dx$, the momentum
is $H_{1}=\int \rho udx$ and the Casimir is $H_{0}=\int \rho dx$.
Corresponding Lagrangian representation is%
\begin{equation}
S_{1}=\int [\frac{z_{t}^{2}}{2z_{x}}+\mu ^{\prime \prime }(z)]dxdt,
\label{dev}
\end{equation}%
where $u=-z_{t}/z_{x}$. Thus, the Lagrangian%
\begin{equation*}
S_{1}=\int [\frac{\Phi _{xt}^{2}}{2\Phi _{xx}}+\mu ^{\prime \prime }(\Phi
_{x})]dxdt
\end{equation*}%
creates the Euler-Lagrange equation%
\begin{equation}
\Phi _{xx}\Phi _{tt}-\Phi _{xt}^{2}=\mu ^{\prime \prime }(\Phi _{x})\Phi
_{xx},  \label{elev}
\end{equation}%
which is a Monge-Ampere equation (cf. (\ref{mon})).

At the same time (\ref{elev}) allows other Lagrangian representation (see
\cite{ns})
\begin{equation}
S_{2}=\int [\frac{1}{2}\Phi _{xx}\Phi _{t}^{2}-\mu (\Phi _{x})]dxdt.
\label{int}
\end{equation}%
Thus, the Monge-Ampere equation (\ref{elev}) has the second Hamiltonian
structure%
\begin{equation*}
r_{t}=\partial _{x}\frac{\delta \bar{H}_{2}}{\delta z}\text{, \ \ \ \ \ }%
z_{t}=\partial _{x}\frac{\delta \bar{H}_{2}}{\delta r}
\end{equation*}%
determined by the local Poisson bracket%
\begin{equation}
\{z(x)\text{, }r(x^{\prime })\}_{2}=\{r(x)\text{, }z(x^{\prime
})\}_{2}=\delta ^{\prime }(x-x^{\prime }),  \label{dozen}
\end{equation}%
where $r=\Phi _{xx}\Phi _{t}$, the Hamiltonian is $\bar{H}_{2}=\int [\frac{%
r^{2}}{2z_{x}}+\mu (z)]dx$, the momentum is $\bar{H}_{1}=\int rzdx$, two
Casimirs are $\bar{Q}_{1}=\int rdx$ and $\bar{Q}_{2}=\int zdx$. Four local
conservation laws associate with above second Hamiltonian structure are
\begin{eqnarray*}
z_{t} &=&\partial _{x}(\frac{r}{z_{x}}),\text{ \ \ \ \ \ \ \ \ \ \ \ \ \ \ \
\ \ \ }\partial _{t}[\frac{r^{2}}{2z_{x}}+\mu (z)]=\partial _{x}[\mu
^{\prime }(z)\frac{r}{z_{x}}+\frac{1}{6}\partial _{x}(\frac{r^{3}}{z_{x}^{3}}%
)], \\
r_{t} &=&\partial _{x}[\mu ^{\prime }(z)+\partial _{x}(\frac{r^{2}}{%
2z_{x}^{2}})]\text{, \ \ \ \ \ \ }\partial _{t}(rz)=\partial _{x}[z\mu
^{\prime }(z)-\mu (z)+\frac{z}{2}\partial _{x}(\frac{r^{2}}{z_{x}^{2}})].
\end{eqnarray*}

\section{Recursion Operator. Integrability of the GZ hierarchy}

Applying the reciprocal transformation (\ref{7}) simultaneously to both
Lagrangian representations (\ref{dev}) and (\ref{int}), one obtains the
variation principles in \textit{another} independent variables%
\begin{equation*}
S_{1}=\int [\frac{1}{2}x_{\tau }^{2}+\mu ^{\prime \prime
}(z)x_{z}]dzd\tau\ \ \ \ \ \text{and \ }\ \ \ S_{2}=\int
[\frac{1}{2}\tilde{\Phi}_{\tau }^{2}-\mu
(z)\tilde{\Phi}_{zz}]dzd\tau,
\end{equation*}%
where $u=x_{\tau }$, \ $\rho ^{-1}=x_{z}$,\ \ $x=\tilde{\Phi}_{z}$, (see the
first conservation law associated with the second Hamiltonian structure (\ref{dozen}): $%
d\Phi =zdx+\frac{r}{\rho }dt$ or $d\tilde{\Phi}=xdz-\frac{r}{\rho
}d\tau $, where $xz=\Phi +\tilde{\Phi}$). Then the Euler-Lagrange
equation is $x_{\tau \tau }=-\mu ^{\prime \prime \prime }(z)$.
This equation can be easily
integrated (see (\ref{9})). Corresponding Poisson brackets (see (\ref{eight}%
) and (\ref{dozen}))%
\begin{eqnarray*}
\{x(z)\text{, }u(z^{\prime })\}_{1} &=&-\{u(z)\text{, }x(z^{\prime
})\}_{1}=\delta (z-z^{\prime }), \\
\{p(z)\text{, }\tilde{\Phi}(z^{\prime })\}_{2}
&=&-\{\tilde{\Phi}(z)\text{, }p(z^{\prime })\}_{2}=\delta
(z-z^{\prime }),
\end{eqnarray*}%
where $u=\tilde{\Phi}_{z\tau }$ and $d\tilde{\Phi}=xdz-pd\tau $ (i.e. $%
p=r/\rho\equiv \Phi _{t}$), create the recursion operator%
\begin{equation*}
\hat{R}=-\left(
\begin{tabular}{ll}
$\partial _{z}^{2}$ &  \\
& $\partial _{z}^{2}$%
\end{tabular}%
\ \right)
\end{equation*}%
where%
\begin{equation*}
\{x(z)\text{, }u(z^{\prime })\}_{2}=-\{u(z)\text{, }x(z^{\prime
})\}_{2}=-\delta ^{\prime \prime }(z-z^{\prime }).
\end{equation*}%
Thus, the Gurevich-Zybin system in these independent variables has an
\textit{infinite set of \textbf{local} Hamiltonian structures, conservation
laws and commuting flows}. For instance, all such Hamiltonians are%
\begin{equation*}
\tilde{H}_{k}=(-1)^{k}\int [\frac{1}{2}p^{(k)^{2}}+\mu ^{(k+2)}(z)\tilde{\Phi%
}^{(k)}]dz\text{, \ \ \ \ \ }k=0\text{, }\pm 1\text{, }\pm 2\text{, ...}
\end{equation*}%
Corresponding commuting flows%
\begin{equation*}
\tilde{\Phi}_{\tau ^{k}\tau ^{k}}=-\mu ^{(k+2)}(z)
\end{equation*}%
can be easily integrated (see (\ref{9})). However, in the independent
variables ($x$, $t^{k}$) they can be written in the form (cf. (\ref{ex}))%
\begin{equation}
\partial _{t^{k}}u^{k}+u^{k}\partial _{x}u^{k}+\mu ^{(2k+3)}(z)=0\text{, \ \
\ \ \ }\partial _{t^{k}}z+u^{k}\partial _{x}z=0,  \label{fif}
\end{equation}%
where%
\begin{equation}
u^{0}\equiv u\text{, \ \ \ \ }u^{k+1}=\frac{1}{\rho }\partial _{x}u^{k}\text{%
, \ \ \ \ \ }u^{-k-1}=\partial _{x}^{-1}(\rho u^{-k})\text{, \ \ \ \ \ }k=0%
\text{, }1\text{, }2\text{, ...}  \label{sh}
\end{equation}%
Thus, \textit{all commuting flows} to the Gurevich-Zybin system created by
above bi-Hamiltonian structure are \textit{Monge-Ampere equations (cf. (\ref%
{elev}))}%
\begin{equation*}
\Phi _{xx}\Phi _{t^{k}t^{k}}-\Phi _{xt^{k}}^{2}=\mu ^{(2k+2)}(\Phi _{x})\Phi
_{xx},
\end{equation*}%
where%
\begin{equation*}
u^{k}=-\Phi _{t^{k+1}}\text{, \ \ \ \ \ \ }\Phi _{xt^{k}}=\Phi _{xx}\Phi
_{t^{k+1}}\text{, \ \ \ \ \ }k=0\text{, }\pm 1\text{, }\pm 2\text{, ...}
\end{equation*}%
All local Lagrangians are%
\begin{eqnarray*}
S_{2,k} &=&\int [\frac{1}{2}\Phi _{xx}\Phi _{t^{k}}^{2}-\mu ^{(2k)}(\Phi
_{x})]dxdt^{k}\text{,} \\
S_{1,k} &=&\int [\frac{\Phi _{xt^{k}}^{2}}{2\Phi _{xx}}+\mu ^{(2k+2)}(\Phi
_{x})]dxdt^{k}\text{,} \\
S_{0,k} &=&\int [\frac{1}{2\Phi _{xx}}[(\frac{\Phi _{xt^{k}}}{\Phi _{xx}}%
)_{x}]^{2}-\mu ^{(2k+4)}(\Phi _{x})]dxdt^{k}, \\
S_{-1,k} &=&\int [\frac{1}{2\Phi _{xx}}[(\frac{1}{\Phi _{xx}}[(\frac{\Phi
_{xt^{k}}}{\Phi _{xx}})_{x}])_{x}]^{2}+\mu ^{(2k+6)}(\Phi _{x})]dxdt^{k}%
\text{, ...}
\end{eqnarray*}%
\textbf{Remark 2}: All commuting flows have infinitely many different \textit{%
local} representations via different pairs of field variables ($z$, $u^{k}$%
), see (\ref{sh}). For instance (cf. (\ref{fif}))%
\begin{equation*}
z_{t^{k}}+\partial _{x}u^{k-1}=0\text{, \ \ \ \ \ \ }\partial
_{t^{k}}u^{k-1}+\frac{(u_{x}^{k-1})^{2}}{z_{x}}+\mu ^{(2k+2)}(z)=0,
\end{equation*}%
\begin{equation*}
z_{t^{k}}+\partial _{x}(\frac{u_{x}^{k-2}}{z_{x}})=0,\ \ \ \ \ \ \ \partial
_{t^{k}}u^{k-2}+\partial _{x}[\frac{(u_{x}^{k-2})^{2}}{2z_{x}^{2}}]+\mu
^{(2k+1)}(z)=0.
\end{equation*}%
The theory of integrable systems with a multi-Lagrangian structure is
presented in \cite{mpnu}. Usually, every local Lagrangian creates a nonlocal
Hamiltonian structure. Such explicit formulas of nonlocal Hamiltonian
structures, nonlocal commuting flows, nonlocal conservation laws as well as
nonlocal Lagrangians can be found iteratively from already given above
formulas.

\section{Another bi-Hamiltonian structure}

Now in this and two next sections we identify $\upsilon \equiv z$, i.e. we
concentrate attention on case $\mu ^{\prime \prime \prime }(z)=z$, see (\ref%
{4})). In two previous sections we discuss bi-Hamiltonian structure of the
Gurevich-Zybin system. Here we preserve the first Hamiltonian structure (see
(\ref{sev}), (\ref{fir})), but change the second one! In above section we
proved that the Gurevich-Zybin system has infinitely many local Hamiltonian
structures and Lagrangian representations (a general theory is presented in
\cite{mpnu}, see also \cite{nutku}). However, this new second Hamiltonian
structure (see below) is \textbf{not} from this set!

The Gurevich-Zybin system (\ref{3}) is an Euler-Lagrange equation of
corresponding variational principle (see (\ref{dev}), when $\mu ^{\prime
\prime }(z)=z^{2}/2$)%
\begin{equation*}
S=\frac{1}{2}\int [\frac{z_{t}^{2}}{z_{x}}+z^{2}]dxdt.
\end{equation*}

However, the \textit{astonished} fact is that the\textit{\ Gurevich-Zybin
system (\ref{3}) has \textbf{another} Hamiltonian structure connected with
the \textbf{same} Lagrangian density}. Namely (see for details \cite{nutku},
especially formulas (\textbf{43)}, (\textbf{52}-\textbf{54)} therein), the
Lagrangian (cf. with $S$)%
\begin{equation*}
\tilde{S}=\frac{1}{2}\int [\frac{p_{x}}{z_{x}}(2z_{t}-p_{x})+z^{2}]dxdt
\end{equation*}%
determines the same Euler-Lagrange equations (\ref{3}) but with another
Hamiltonian structure%
\begin{equation*}
u_{t}=-\partial _{x}^{-1}\frac{\delta H_{1}}{\delta u}+u_{x}\partial
_{x}^{-1}\frac{\delta H_{1}}{\delta z}\text{, \ \ \ \ \ }z_{t}=\partial
_{x}^{-1}(u_{x}\frac{\delta H_{1}}{\delta u}+z_{x}\frac{\delta H_{1}}{\delta
z})+z_{x}\partial _{x}^{-1}\frac{\delta H_{1}}{\delta z},
\end{equation*}%
where $u=-p_{x}/z_{x}$ (i.e. $p=\Phi _{t}$).

\textbf{Remark 3}: This bi-Hamiltonian structure at first was
discovered by Yavuz Nutku \cite{yavuz} and later it was
independently found in \cite{das} (see formula (\textbf{9})
therein) exactly as it was done in \cite{nutku}. However, here we
repeat and emphasize the \textbf{main observation} of this
section is that \textit{\textbf{both} Hamiltonian structures have the \textbf{%
same} Lagrangian density}! This is the first such example in the
theory of integrable systems.

\subsection*{Canonical representation for both \textit{Hamiltonian }%
structures and recursion operator}

The Poisson bracket%
\begin{eqnarray*}
\{u(x)\text{, }u(x^{\prime })\}_{1} &=&0\text{, \ \ \ \ \ \ }\{\rho (x)\text{%
, }u(x^{\prime })\}_{1}=\delta ^{\prime }(x-x^{\prime }), \\
\{u(x)\text{, }\rho (x^{\prime })\}_{1} &=&\delta ^{\prime }(x-x^{\prime })%
\text{, \ \ \ \ \ \ }\{\rho (x)\text{, }\rho (x^{\prime })\}_{1}=0
\end{eqnarray*}%
of the first Hamiltonian structure is given in its canonical form (more
details see in the review \cite{dubr}). However, the Poisson bracket
\begin{eqnarray*}
\{u(x)\text{, }u(x^{\prime })\}_{2} &=&-\partial ^{-1}\delta (x-x^{\prime })%
\text{, \ \ \ \ \ }\{u(x)\text{, }\rho (x^{\prime })\}_{2}=-u_{x}\delta
(x-x^{\prime }), \\
\{\rho (x)\text{, }u(x^{\prime })\}_{2} &=&u_{x}\delta (x-x^{\prime })\text{%
, \ \ \ }\{\rho (x)\text{, }\rho (x^{\prime })\}_{2}=-(\rho \partial
_{x}+\partial _{x}\rho )\delta (x-x^{\prime })
\end{eqnarray*}%
of the second Hamiltonian structure can be reduced by the Darboux theorem to
the canonical form%
\begin{eqnarray*}
\{w(x)\text{, }w(x^{\prime })\}_{2} &=&\delta ^{\prime }(x-x^{\prime })\text{%
, \ \ \ \ \ }\{\eta (x)\text{, }w(x^{\prime })\}_{2}=0, \\
\{w(x)\text{, }\eta (x^{\prime })\}_{2} &=&0\text{, \ \ \ \ \ \ \ }\{\eta (x)%
\text{, }\eta (x^{\prime })\}_{2}=\delta ^{\prime }(x-x^{\prime })
\end{eqnarray*}%
by the Miura type transformation (see the first equation in (\ref{ra}))%
\begin{equation*}
w=u_{x}\text{, \ \ \ \ \ \ }\rho =-\frac{1}{2}(w^{2}+\eta ^{2}).
\end{equation*}%
Then the Gurevich-Zybin system written in a \textit{modified} form (see
formula (\textbf{26}) in \cite{mpl}, other details in the last section
\textbf{5} and references [\textbf{6}], [\textbf{7}] therein)%
\begin{equation}
\eta _{t}+\partial _{x}(u\eta )=0\text{, \ \ \ \ \ \ }u_{xt}+uu_{xx}+\frac{1%
}{2}u_{x}^{2}=\frac{1}{2}\eta ^{2}  \label{moda}
\end{equation}%
can be recognized as the two-component generalization of the Hunter-Saxton
equation (cf. \cite{hs}, \cite{dp}, \cite{or}).

\textbf{Remark 4}: In fact, the Casimir density $\eta $ of the
second Hamiltonian structure was found in \cite{das} (see formula
(\textbf{18})
therein). However, the Gurevich-Zybin system was not presented in form (\ref%
{moda}) there. Moreover, we emphasize the \textbf{main result} of this paper
is that \textit{the Gurevich-Zybin system belongs to the well-known class of
integrable systems.} In this section we prove that the Gurevich-Zybin system
is a member of an integrable hierarchy embedded into 2x2 spectral transform.

Since the first Poisson bracket in new field variables has a form%
\begin{eqnarray*}
\{w(x),\eta (x^{\prime })\}_{1} &=&-[\frac{1}{\eta }\delta (x-x^{\prime
})]^{\prime \prime },\text{ \ \ \ \ \ \ \ \ }\{\eta (x),w(x^{\prime })\}_{1}=%
\frac{1}{\eta }\delta ^{\prime \prime }(x-x^{\prime }), \\
\{w(x),w(x^{\prime })\}_{1} &=&0\text{, \ \ \ \ \ \ \ \ }\{\eta (x),\eta
(x^{\prime })\}_{1}=-[\frac{w_{x}}{\eta ^{2}}\partial _{x}+\partial _{x}%
\frac{w_{x}}{\eta ^{2}}]\delta (x-x^{\prime }),
\end{eqnarray*}%
then the modified Gurevich-Zybin system (\ref{moda}) as a member of
integrable hierarchy with all other commuting flows together can be written
in bi-Hamiltonian form%
\begin{eqnarray*}
w_{t^{k}} &=&\partial _{x}\frac{\delta H_{k+1}}{\delta w}=-\partial _{x}^{2}[%
\frac{1}{\eta }\cdot \frac{\delta H_{k}}{\delta \eta }], \\
\eta _{t^{k}} &=&\partial _{x}\frac{\delta H_{k+1}}{\delta \eta }=\frac{1}{%
\eta }\partial _{x}^{2}\frac{\delta H_{k}}{\delta w}-[2\frac{w_{x}}{\eta ^{2}%
}\partial _{x}+(\frac{w_{x}}{\eta ^{2}})_{x}]\frac{\delta H_{k}}{\delta \eta
},
\end{eqnarray*}%
An eigenvalue problem for the recursion operator as a ratio of both
Hamiltonian structures
\begin{equation*}
\left[
\begin{array}{ll}
0 & -\partial _{x}^{2}\frac{1}{\eta } \\
\frac{1}{\eta }\partial _{x}^{2} & -(\frac{w_{x}}{\eta ^{2}}\partial
_{x}+\partial _{x}\frac{w_{x}}{\eta ^{2}})%
\end{array}%
\right] (%
\begin{array}{l}
\varphi _{1} \\
\varphi _{2}%
\end{array}%
)=2\lambda \partial _{x}(%
\begin{array}{l}
\varphi _{1} \\
\varphi _{2}%
\end{array}%
).
\end{equation*}%
can be written as one equation%
\begin{equation*}
\varphi _{xxx}+4(\lambda ^{2}\eta ^{2}+\lambda \sigma )\varphi
_{x}+2(\lambda ^{2}\eta ^{2}+\lambda \sigma )_{x}\varphi =0,
\end{equation*}%
where $\varphi _{1}=\varphi _{x}$, \ $\varphi _{2}=-2\lambda \eta \varphi $ and $%
\sigma =w_{x}$. However, above equation can be reduced to%
\begin{equation*}
\psi _{xx}+(\lambda ^{2}\eta ^{2}+\lambda \sigma )\psi =0,
\end{equation*}%
where $\varphi =\psi \psi ^{+}$ is a \textit{squared
eigenfunction} and $\psi $, $\psi ^{+}$ are linear conjugate
solutions with different asymptotics at infinity $\lambda
\rightarrow \infty $. This linear spectral problem (more
precisely, just \textquotedblleft $x$\textquotedblright
--dynamics) is well known in the theory of integrable systems:
corresponding systems are members (commuting
flows) of the two-component Harry Dym hierarchy (see for instance \cite{af}%
). All such members of this hierarchy can be determined by the spectral
transform
\begin{equation}
\psi _{xx}=-(\lambda ^{2}\eta ^{2}+\lambda \sigma )\psi \text{, \ \ \ \ \ \ }%
\psi _{t}=b\psi _{x}-\frac{1}{2}b_{x}\psi ,  \label{dyna}
\end{equation}%
where $b(\zeta ,\eta ,\lambda )$ is a \textit{polynomial} function with
respect to the spectral parameter $\lambda $ for \textit{positive} members.
The compatibility condition $(\psi _{xx})_{t}=(\psi _{t})_{xx}$ yields
relationship
\begin{equation*}
(\lambda ^{2}\eta ^{2}+\lambda \sigma )_{t}=[\frac{1}{2}\partial
_{x}^{3}+2(\lambda ^{2}\eta ^{2}+\lambda \sigma )\partial _{x}+(\lambda
^{2}\eta ^{2}+\lambda \sigma )_{x}]b,
\end{equation*}%
where the two-component Harry Dym system (see \cite{af})%
\begin{equation*}
\eta _{t_{1}}=(\frac{\sigma }{\eta ^{2}})_{x}\text{, \ \ }\sigma _{t_{1}}=(%
\frac{1}{\eta })_{xxx}
\end{equation*}%
can be obtained if $b=2\lambda /\eta $. Thus, \textit{twice potential}
two-component Harry Dym system%
\begin{equation}
\eta _{t_{1}}=(\frac{u_{xx}}{\eta ^{2}})_{x}\text{, \ \ }u_{t_{1}}=(\frac{1}{%
\eta })_{x}  \label{hd}
\end{equation}%
is the \textit{first} member of \textit{positive} part of above hierarchy
and the \textit{first} member of its \textit{negative} part is the \textit{%
modified} Gurevich-Zybin system (\ref{moda})%
\begin{equation}
\eta _{t_{-1}}+\partial _{x}(u\eta )=0\text{, \ \ \ \ \ \ }%
u_{xt_{-1}}+uu_{xx}+\frac{1}{2}u_{x}^{2}=\frac{1}{2}\eta ^{2}  \label{mgz}
\end{equation}%
determined by the choice $b=(2\lambda )^{-1}-u$ (it means that we \textbf{%
must} identify $t\equiv t_{-1}$ for the Gurevich-Zybin system (\ref{3})).

\textbf{Remark 5}: The reciprocal transformation (see (\ref{7}))%
\begin{equation*}
d\tau _{1}=dt_{1}\text{, \ \ \ \ \ \ }d\tau _{-1}=dt_{-1}\text{, \ \ \ \ \ }%
dz=\rho dx-\rho udt_{-1}-(\frac{u_{x}}{\eta })_{x}dt_{1}
\end{equation*}%
simultaneously linearizes the Gurevich-Zybin system (see (\ref{8}) and (\ref{9})) and \textit{preserves} two-component Harry Dym system:%
\begin{equation*}
\rho _{t_{1}}=-(\frac{w}{\eta })_{xx}\text{, \ \ }w_{t_{1}}=(\frac{1}{\eta }%
)_{xx}\text{, \ \ }\eta _{t_{1}}=(\frac{w_{x}}{\eta ^{2}})_{x}\text{\ \ }%
\rightarrow \text{\ \ }\bar{\rho}_{\tau _{1}}=-(\frac{\bar{w}}{\bar{\eta}}%
)_{zz}\text{, \ \ }\bar{w}_{\tau
_{1}}=(\frac{1}{\bar{\eta}})_{zz}\text{, \ \ }\bar{\eta}_{\tau
_{1}}=(\frac{\bar{w}_{z}}{\bar{\eta}^{2}})_{z},
\end{equation*}%
where%
\begin{equation*}
\bar{\rho}=\frac{1}{\rho }\text{, \ \ \ \ \
}\bar{w}=-\frac{w}{\rho }\text{, \ \ \ \ \ }\bar{\eta}=\frac{\eta
}{\rho }.
\end{equation*}%
Such reciprocal \textit{auto-transformation} is the first example
in the theory of integrable systems.

\textbf{Remark 6}: Twice potential two-component Harry Dym system
(\ref{hd}) written in field variables ($\rho $, $u$) was also
found in \cite{das} (see formula (\textbf{21}) therein), but was
not \textit{recognized}.

\section{Reciprocal and Miura type transformations}

Application of the reciprocal transformation (in fact, it was given in \cite%
{af}, formulas (\textbf{32-34)} therein; cf. (\ref{eta}))%
\begin{equation}
dy_{1}=dt_{1}\text{, \ \ \ \ \ \ }dy_{-1}=dt_{-1}\text{, \ \ \ \ \
}d\zeta =\eta dx-\eta udt_{-1}+\frac{u_{xx}}{\eta ^{2}}dt_{1}
\label{rt}
\end{equation}%
to the spectral transform (\ref{dyna})\ yields another well-known spectral
transform (see for instance, \cite{af}, \cite{mpcc}), where
\textquotedblleft $\zeta $\textquotedblright --dynamics is%
\begin{equation}
\tilde{\psi} _{\zeta \zeta }+[\lambda ^{2}-\tilde{u}\lambda -\tilde{\upsilon}+%
\frac{\tilde{u}^{2}}{4}]\tilde{\psi} =0,  \label{tri}
\end{equation}%
\textquotedblleft $y$\textquotedblright --dynamics is%
\begin{equation}
\tilde{\psi} _{y_{1}}=(2\lambda +\tilde{u})\tilde{\psi} _{\zeta }-\frac{1}{2}\tilde{u}%
_{\zeta }\tilde{\psi} \text{, \ \ \ \ \ \ \ \ }\tilde{\psi} _{y_{-1}}=\frac{1}{%
4\lambda }(2\eta \tilde{\psi} _{\zeta }-\eta _{\zeta }\tilde{\psi}
) \label{chet}
\end{equation}%
and%
\begin{equation}
\tilde{\psi} =\eta ^{1/2}\psi \text{, \ \ \ \ \ \ }-\tilde{u}=u_{\zeta \zeta }+%
\frac{\eta _{\zeta }}{\eta }u_{\zeta }\text{, \ \ \ \ \ }-\tilde{\upsilon}+%
\frac{\tilde{u}^{2}}{4}=\frac{\eta _{\zeta }^{2}}{4\eta ^{2}}-\frac{\eta
_{\zeta \zeta }}{2\eta }.  \label{pat}
\end{equation}%
The compatibility conditions $(\tilde{\psi} _{\zeta \zeta
})_{y_{1}}=(\tilde{\psi} _{y_{1}})_{\zeta \zeta }$ and
$(\tilde{\psi} _{\zeta \zeta })_{y_{-1}}=(\tilde{\psi}
_{y_{-1}})_{\zeta \zeta }$ yield the first \textit{positive} member%
\begin{equation}
\tilde{u}_{y_{1}}=2\partial _{\zeta }[\frac{\tilde{u}^{2}}{2}+\tilde{\upsilon%
}]\text{, \ \ \ \ \ }\tilde{\upsilon}_{y_{1}}=2\partial _{\zeta }[\tilde{u}%
\tilde{\upsilon}-\frac{1}{4}\tilde{u}_{\zeta \zeta }],  \label{kb}
\end{equation}%
and the first \textit{negative} member%
\begin{equation*}
\tilde{u}_{y_{-1}}=-\eta _{\zeta }\text{, \ \ \ \ \ }\tilde{\upsilon}%
_{y_{-1}}=\frac{1}{2}\partial _{\zeta }(\tilde{u}\eta )\text{, \ \ \ \ \ }-%
\frac{1}{2}\eta _{\zeta \zeta \zeta }+(2\tilde{\upsilon}-\frac{\tilde{u}^{2}%
}{2})\eta _{\zeta }+(\tilde{\upsilon}_{\zeta }-\frac{1}{2}\tilde{u}\tilde{u}%
_{\zeta })\eta =0
\end{equation*}%
which also can be obtain by the limit $\tilde{\lambda}\rightarrow 0$ from
the generating function of commuting flows%
\begin{equation*}
\tilde{u}_{y}=-\tilde{\eta}_{\zeta }\text{, \ \ \ }\tilde{\upsilon}%
_{y}=\partial _{\zeta }[(\frac{1}{2}\tilde{u}-\tilde{\lambda})\tilde{\eta}]%
\text{,}
\end{equation*}%
\begin{equation}
\tilde{\eta}_{\zeta \zeta \zeta }+4(\tilde{\lambda}^{2}-\tilde{u}\tilde{%
\lambda}-\tilde{\upsilon}+\frac{\tilde{u}^{2}}{4})\tilde{\eta}_{\zeta }+2(-%
\tilde{\lambda}\tilde{u}_{\zeta }-\tilde{\upsilon}_{\zeta }+\frac{1}{2}%
\tilde{u}\tilde{u}_{\zeta })\tilde{\eta}=0,  \label{gen}
\end{equation}%
where $\tilde{\eta}=\tilde{\varphi}\tilde{\varphi}^{+}$ is a \textit{squared eigenfunction} and $\tilde{\varphi}$%
, $\tilde{\varphi}^{+}$ are linear conjugate solutions of the
spectral
transform (\ref{tri}) with different asymptotics at the infinity $\tilde{%
\lambda}\rightarrow \infty $ (see details for instance in \cite{mpwh}).

At the same time, as it was proved in \cite{mpl} (see also above section
\textbf{3}), the modified Gurevich-Zybin system is \textit{linearized} by
above reciprocal transformation (\ref{rt}). Simultaneously, the \textit{%
twice potential} two-component Harry Dym system (\ref{hd}) transforms into
\textit{twice degenerate twice modified} Kaup-Boussinesq system%
\begin{equation}
u_{y_{1}}=-u_{\zeta }u_{\zeta \zeta }-\frac{\eta _{\zeta }}{\eta }%
(1+u_{\zeta }^{2})\text{, \ \ \ \ \ }\eta _{y_{1}}=\eta u_{\zeta \zeta \zeta
}+(\eta _{\zeta \zeta }-2\frac{\eta _{\zeta }^{2}}{\eta })u_{\zeta }.
\label{k}
\end{equation}%
It is well-known that (\ref{kb}) is the Kaup-Boussinesq system (see for
instance \cite{kb}). Several modified Kaup-Boussinesq systems were presented
in \cite{mpcc}. The \textit{modified} Kaup-Boussinesq system is%
\begin{equation*}
\tilde{u}_{y_{1}}=\partial _{\zeta }[\frac{3}{2}\tilde{u}%
^{2}+2u_{1}^{2}+2u_{1,\zeta }]\text{, \ \ \ \ \ }u_{1,y_{1}}=\partial
_{\zeta }[u_{1}\tilde{u}-\frac{1}{2}\tilde{u}_{\zeta }],
\end{equation*}%
where the \textit{first} Miura transformation is%
\begin{equation}
\tilde{\upsilon}=\frac{1}{4}\tilde{u}^{2}+u_{1}^{2}+u_{1,\zeta }.
\label{miura}
\end{equation}%
The \textit{twice modified} Kaup-Boussinesq system is%
\begin{equation*}
u_{1,y_{1}}=\partial _{\zeta }[(2u_{1}^{2}-u_{1,\zeta })u_{2}-\frac{1}{2}%
u_{2,\zeta \zeta }]\text{, \ \ \ \ \ }u_{2,y_{1}}=\partial _{\zeta
}[2u_{1}(1+u_{2}^{2})+u_{2}u_{2,\zeta }],
\end{equation*}%
where the \textit{second} Miura transformation is%
\begin{equation}
\tilde{u}=2u_{1}u_{2}+u_{2,\zeta }\text{.}  \label{miu}
\end{equation}%
It was proved in \cite{mpcc} that the Kaup-Boussinesq system has third and
fourth Miura transformations (see also \cite{bor}). Their \textit{double}
\textit{parametric degeneration} to \textit{purely potential} form (cf. (\ref%
{miura}), (\ref{miu}) with second and third equations from (\ref{pat}))%
\begin{equation}
u_{1}=\frac{1}{2}\partial _{\zeta }\ln \eta \text{, \ \ \ \ \ \ }%
u_{2}=-u_{\zeta }  \label{pot}
\end{equation}%
transforms the twice modified Kaup-Boussinesq system in form (\ref{k}).

Thus, the \textbf{main result} of this section is an establishment of the
link of transformations (reciprocal and Miura type) between the
Gurevich-Zybin and the Kaup-Boussinesq hierarchies.

\section{Second Harmonic Generation}

The generation of the second harmonic wave from the red light of a ruby
laser in a crystal of quartz in fact was the starting point of nonlinear
optics. In one-dimensional case, for short pulses, when the group-velocity
mismatch between both frequency components becomes important, then the
process of second harmonic generation (SHG) is described by the complex
equations (see for instance \cite{kshg}, \cite{ks}), which in real form are
(see \cite{ks}, formula (\textbf{6}) therein)%
\begin{equation}
\tilde{u}_{y_{-1}}=-\eta _{\zeta }=-2\eta u_{1}\text{, \ \ \ \ \ \ }%
2u_{1,y_{-1}}=-\frac{1}{\eta }+\eta \tilde{u}.  \label{shg}
\end{equation}%
The corresponding spectral problem (see again \cite{ks}, formula (\textbf{9}%
) therein)%
\begin{equation*}
\tilde{\psi} _{\zeta \zeta }+[\lambda ^{2}-\tilde{u}\lambda
-u_{1}^{2}-u_{1,\zeta }]\tilde{\psi} =0\text{, \ \ \ \ \ \ \
}\tilde{\psi} _{y_{-1}}=\frac{1}{2\lambda }\eta \tilde{\psi}
_{\zeta }-\frac{1}{2\lambda }\eta u_{1}\tilde{\psi}
\end{equation*}%
is a \textit{special} reduction of the spectral transform (\ref{tri}), (\ref%
{chet}). In general case (\ref{gen}) can be integrated once%
\begin{equation*}
\tilde{\eta}\tilde{\eta}_{\zeta \zeta }-\frac{1}{2}\tilde{\eta}_{\zeta
}^{2}+2(\tilde{\lambda}^{2}-\tilde{u}\tilde{\lambda}-\tilde{\upsilon}+\frac{%
\tilde{u}^{2}}{4})\tilde{\eta}^{2}+S(\tilde{\lambda})=0,
\end{equation*}%
where $S(\tilde{\lambda})$ is a polynomial function for multi-periodic
solutions of the Kaup-Boussinesq hierarchy (see for instance \cite{alber}).
Thus, the first \textit{negative} member of this hierarchy has the constraint%
\begin{equation*}
\eta \eta _{\zeta \zeta }-\frac{1}{2}\eta _{\zeta }^{2}+2(-\tilde{\upsilon}+%
\frac{\tilde{u}^{2}}{4})\eta ^{2}+S_{-1}=0,
\end{equation*}%
where $S_{-1}\neq 0$ is a some constant. However, $S_{-1}=0$ in the case of
the SHG system! It is easy to prove by direct substitution (\ref{miura}) and
$\eta _{\zeta }=2\eta u_{1}$ from (\ref{shg}) (see also the first equation
in (\ref{pot})) in above equation. Thus, the SHG system (\ref{shg}) is the
\textit{degenerate first negative} member of the \textit{modified}
Kaup-Boussinesq hierarchy (see (\ref{miura}) and above).

The SHG system%
\begin{equation}
(\ln \Xi _{y_{-1}})_{\zeta y_{-1}}=\frac{1}{\Xi _{y_{-1}}}-\Xi _{\zeta }\Xi
_{y_{-1}}  \label{shgful}
\end{equation}%
can be interpreted as the two-component generalization of the Sinh-Gordon
equation, where $\Xi _{\zeta }=\tilde{u}$ and $\Xi _{y_{-1}}=-\eta $. The
SHG system has \textit{three} different linearizable degenerations, as well
as the Sinh-Gordon equation has a \textit{parametric} degeneration to the
famous Liouville equation which is linearizable. First two degenerate limits
are known (see \cite{bass} and \cite{akmp}). These are the Liouville
equation and the modified Liouville equation. The \textit{third} such case
can be obtained by (see above) differential substitutions (\ref{miu}) and (%
\ref{pot}). This is the modified Gurevich-Zybin system (\ref{mgz})
re-calculated by the reciprocal transformation (\ref{rt}) (see formula (%
\textbf{27}) in \cite{mpl} and other details in the last section \textbf{5})%
\begin{equation*}
u_{\zeta }=(\frac{1}{\eta })_{y_{-1}}\text{, \ \ \ \ }\eta _{y_{-1}y_{-1}}-%
\frac{3\eta _{y_{-1}}^{2}}{2\eta }+\frac{1}{2}\eta ^{3}=0.
\end{equation*}%
Thus, a solution of the \textit{reduced} SHG system%
\begin{equation*}
(u_{\zeta \zeta }+\frac{\eta _{\zeta }}{\eta }u_{\zeta })_{y_{-1}}=\eta
_{\zeta }\text{, \ \ \ \ \ \ }(\ln \eta )_{\zeta y_{-1}}=-\frac{1}{\eta }%
-(\eta u_{\zeta })_{\zeta }
\end{equation*}%
in an implicit form is given by (\ref{13})%
\begin{eqnarray*}
\eta  &=&[\frac{1}{\eta _{0}(\zeta )}+\frac{\eta _{0}(\zeta )}{4}%
(y_{-1}-y_{0}(\zeta ))^{2}]^{-1}, \\
u &=&\frac{y_{-1}}{2}\int \eta _{0}(\zeta )d\zeta -\frac{1}{2}\int \eta
_{0}(\zeta )y_{0}(\zeta )d\zeta , \\
x &=&\int [\frac{1}{\eta _{0}(\zeta )}+\eta _{0}(\zeta )y_{0}^{2}(\zeta
)]d\zeta -\frac{y}{2}\int \eta _{0}(\zeta )y_{0}(\zeta )d\zeta +\frac{y^{2}}{%
4}\int \eta _{0}(\zeta )d\zeta .
\end{eqnarray*}%
Thus, a \textit{new} solution of the SHG system (\ref{shgful}) can be found
in quadratures%
\begin{equation*}
d\Xi =-(u_{\zeta \zeta }+\frac{\eta _{\zeta }}{\eta }u_{\zeta })d\zeta -\eta
dy_{-1}.
\end{equation*}%
\textbf{Final remark}: Since the Kaup-Boussinesq system and the nonlinear
Schroedinger equation are related by invertible transformations (see for
instance \cite{alber}), then their first negative flows are related too.
Since the first negative flow to the nonlinear Schroedinger equation is
another famous Maxwell-Bloch system (in particular case the
\textquotedblleft self-induced transparency\textquotedblright\ coincides
with the Maxwell-Bloch system), then the SHG system connected to
Maxwell-Bloch system by the same transformations. Thus, the Gurevich-Zybin
system is connected with the Maxwell-Bloch system. Similarly, a new solution
of the Maxwell-Bloch system can be found by the same way. Since nonlinear
Schroedinger equation relates to Heisenberg magnet by Miura type
transformations, then particular case of the \textquotedblleft Ramann
scattering\textquotedblright\ is connecting with the Maxwell-Bloch system
too. Thus, a new solution for the Ramann scattering can be constructed as in
previous case.

\section{Open Problems}

The numerical simulation of nonlinear dynamics described by the
Gurevich-Zybin system yields the hypothesis that a behaviour in multimode
form demonstrating the transition from the hydrodynamic to the equilibrium
kinetic state has some regular features (see for details \cite{gz}) possibly
generated by \textit{integrable} properties of corresponding $N$--component
Gurevich-Zybin system (see (\ref{2}))%
\begin{equation*}
\rho _{t}^{k}+\partial _{x}(\rho ^{k}u^{k})=0\text{, \ \ \ }%
u_{t}^{k}+u^{k}u_{x}^{k}+\Phi _{x}=0\text{, \ \ \ \ }\Phi _{xx}=\underset{m=1%
}{\overset{N}{\sum }}\rho ^{m}.
\end{equation*}%
This problem (integrability in any sense: linearization, inverse
scattering transform, bi-Hamiltonian formulation, etc) is open.
For instance, this system written in field variables $u^{k}$ and
$\upsilon ^{k}$
($\rho ^{k}\equiv \upsilon _{x}^{k}$) has ultra-local Hamiltonian structure%
\begin{equation*}
\upsilon _{t}^{k}=\frac{\delta H_{2}}{\delta u^{k}}\text{, \ \ \ \ }%
u_{t}^{k}=-\frac{\delta H_{2}}{\delta \upsilon ^{k}},
\end{equation*}%
where the Hamiltonian is%
\begin{equation*}
H_{2}=\frac{1}{2}\int [-\underset{m=1}{\overset{N}{\sum }}%
(u^{m})^{2}\upsilon _{x}^{m}+\left( \underset{m=1}{\overset{N}{\sum }}%
\upsilon ^{m}\right) ^{2}]dx.
\end{equation*}%
It was proved here that the Gurevich-Zybin system has infinitely
many Hamiltonian structures. Existence of the second Hamiltonian
structure is enough for an integrability.

Introducing moments%
\begin{equation}
A_{k}=\underset{m=1}{\overset{N}{\sum }}(u^{m})^{k}\rho ^{m}  \label{14}
\end{equation}%
then $N-$component Gurevich-Zybin system can be written as the nonlocal chain%
\begin{equation}
\partial _{t}A_{k}+\partial _{x}A_{k+1}+kA_{k-1}\partial _{x}^{-1}A_{0}=0%
\text{, \ \ \ \ }k=0\text{, }1\text{, }2\text{, ...,}  \label{15}
\end{equation}%
which looks very similar as famous \textit{integrable} Benney moment chain
(see \cite{Ben})%
\begin{equation*}
\partial _{t}A_{k}+\partial _{x}A_{k+1}+kA_{k-1}\partial _{x}A_{0}=0\text{,
\ \ \ \ }k=0\text{, }1\text{, }2\text{, ...}
\end{equation*}%
The Benney moment chain has infinitely many $N$-component
reductions (see \cite{tsar}) parameterized by $N$ functions of a
single variable ($N$ is an arbitrary natural integer), where the
simplest reduction is (\ref{14}). The nonlocal chain
(\ref{15}) has at least one simple reduction%
\begin{equation*}
A_{k}=\rho u^{k}\text{.}
\end{equation*}%
The existence of any other such reductions could be a
\textit{symptom} of an integrability. The integrability of
$N-$component Gurevich-Zybin system and a description of other
reductions of the nonlocal chain (\ref{15}) will be under
consideration anywhere else.

\section{Conclusion}

The Gurevich-Zybin system is an example of integrable systems
possessing properties of two different classes: $C-$ and $S-$
integrable. This system is linearizable (see \cite{gz}) and has a
general solution. Thus, the Gurevich-Zybin system is from a
$C-$integrable class. However, this system has an infinite set of
Hamiltonian structures and commuting flows. Thus, the
Gurevich-Zybin system is from a $S-$ integrable class too.
Moreover, this system has an infinite set of \textit{local}
Hamiltonian structures, that is unusual in the theory of
$S-$integrable systems. Moreover, all commuting flows of the
Gurevich-Zybin system written in a form of a Monge-Ampere equation
are the same Monge-Ampere equation again. The difference between
them is just some derivative of function $\mu (z)$, which can be
eliminated
by a point transformation $\upsilon =\mu ^{\prime \prime \prime }(z)$ (see (%
\ref{ex})). Thus, this is a beautiful example from mathematical point of
view having physical application (see \cite{gz}). We repeat again: \textit{%
every commuting flow} can be written in the \textit{unique form }(\ref{elev})%
\begin{equation*}
\Phi _{xx}\Phi _{t^{k}t^{k}}-\Phi _{xt^{k}}^{2}=K(\Phi _{x})\Phi _{xx},
\end{equation*}%
where $K(z)$ is \textbf{any} apriori given function, but all commuting flows
will be written via different functions $\Phi _{(k)}$, because the point
transformation like $\upsilon =\mu ^{\prime \prime \prime }(z)$ becames
nonlocal%
\begin{equation*}
K(\Phi _{(k)x})=\mu ^{(2k+2)}(\Phi _{x}).
\end{equation*}%
The exceptional case is when some derivative $\mu ^{(n)}(z)$ is a constant:
then \textquotedblleft half\textquotedblright\ of commuting flows are
trivial (see \cite{nutku}, \cite{fer})%
\begin{equation*}
\Phi _{xx}\Phi _{t^{k}t^{k}}-\Phi _{xt^{k}}^{2}=0\text{,}
\end{equation*}%
when $n$ is even, then\ $k\geqslant n/2$, when $n$ is odd, then\ $k\geqslant
(n-1)/2$.

At past 10 years a couple of such examples of integrable systems (mixed
properties of $C-$ and $S-$ integrability) was found in \cite{nutku} and
\cite{ferlie}. However, an explanation of such phenomenon is not exist to
this moment. One possible explanation is that such systems are in \textit{%
intersection} of $C-$ and $S-$ integrability. Thus, they
\textit{accumulate} properties of these two different classes.
Moreover, we proved that the Gurevich-Zybin system is a
\textit{degenerate} member of the two-component Harry Dym
hierarchy. A \textit{degeneracy} becomes when equations can
possess a \textit{parametric} freedom. When some of parameters are
fixed (to zero, for instance), then such equations become to
\textit{linearizable}.
The simplest example is the famous Liouville equation%
\begin{equation*}
w_{xt}=e^{w}.
\end{equation*}%
This equation is in \textit{intersection} of two different integrable
hierarchies. One of them is another famous Bonnet equation (well known in
physics as the Sinh-Gordon equation) first introduced in a differential
geometry of surfaces of a constant curvature%
\begin{equation*}
w_{xt}=c_{1}e^{w}+c_{2}e^{-w},
\end{equation*}%
which is a member of the potential \textit{modified} KdV hierarchy (spectral
transform 2x2)%
\begin{equation*}
w_{\tau }=w_{xxx}-\frac{1}{2}w_{x}^{3}.
\end{equation*}%
Another one is the Tzitzeica equation, well known in an affine differential
geometry%
\begin{equation*}
w_{xt}=c_{1}e^{w}+c_{2}e^{-2w},
\end{equation*}%
which is a member of the potential \textit{modified} Sawada-Kotera hierarchy
(spectral transform 3x3)%
\begin{equation*}
w_{\tau
}=w_{xxxxx}+5(w_{xx}w_{xxx}-w_{x}^{2}w_{xxx}-w_{x}w_{xx}^{2})+w_{x}^{5}.
\end{equation*}%
Thus, if $c_{2}=0$, then the Liouville equation is still \textit{a member of
two different integrable hierarchies simultaneously}. This is a good \textit{%
symptom} that such equations should be linearizable. Since above mentioned
linearizable reduction of the SHG system is determined by 2x2 spectral
transform, but the SHG system is a some reduction of another important
three-wave interaction problem (see for instance \cite{ks}), then we can
assume such systems like the Gurevich-Zybin system are linearizable if they
are in intersection of at least two different integrable hierarchies (e.g.
the Liouville equation is a member of two different hierarchies: of the KdV
equation and of the Sawada-Kotera equation). A bi-Hamiltonian structures
presented here has origin in 2x2 spectral transform. It will be interesting
to find another Hamiltonian structures coming from 3x3 spectral problem.

Finally, we would like to emphasize that this paper was devoted to \textbf{%
recognition} of a relationship between couple of remarkable systems having
applications in astrophysics, nonlinear optics and geometry.

\section*{Acknowledgments}

I am indebted to Professor Yavuz Nutku for his hospitality (\textbf{1996})
at TUBITAK Marmara Research Centre and in the Feza Gursey Institute
(Istanbul), for his helpful explanation of the relationship between \textit{%
Monge-Ampere }equation and its \textit{Hamiltonian }structures.

I am grateful to Professor Alexander Gurevich for his explanation of the
physical nature of these equations and Prof. Kirill Zybin for his remark
that their system describing a dynamics of dark matter in the Universe can
be derived from the Vlasov kinetic equation (which can be derived from the
Liouville equation for the distribution function).

I would like to thank Dr. Eugene Ferapontov and Dr. Gennady El for their
interest and discussions.

I also wished to thank Professor Andrea Donato for his stimulated talk at
the \textquotedblleft Lie group Analysis\textquotedblright\ conference about
the relationship between some real hydrodynamic type systems and \textit{%
Monge-Ampere }equations. However, few years ago he passed away.

\end{document}